\documentclass{article}

\usepackage{arXiv}

\usepackage[utf8]{inputenc} 
\usepackage[T1]{fontenc}    
\usepackage{hyperref}       
\usepackage{url}            
\usepackage{booktabs}       
\usepackage{amsfonts}       
\usepackage{nicefrac}       
\usepackage{microtype}      
\usepackage{lipsum}		
\usepackage{amsmath, amssymb}
\usepackage{multicol}
\usepackage{multirow}
\usepackage{caption}
\usepackage{xcolor}
\usepackage{blindtext}
\usepackage{graphicx}
\usepackage{doi}

\usepackage[affil-it]{authblk}

\title{Ultrasound Image Enhancement using CycleGAN and Perceptual Loss}

\author[1]{Shreeram Athreya}
\author[2]{Ashwath Radhachandran}
\author[3]{Vedrana Ivezić}
\author[4]{Vivek Sant}
\author[1,2,3,5,6]{Corey W. Arnold}
\author[* 2,3,5]{William Speier}

\affil[1]{Department of Electrical and Computer Engineering, UCLA}
\affil[2]{Department of Bioengineering, UCLA}
\affil[3]{Medical Informatics, UCLA}
\affil[4]{Department of Surgery, UT Southwestern Medical Center}
\affil[5]{Department of Radiological Sciences, UCLA}
\affil[6]{Department of Pathology, UCLA \authorcr Email: \texttt{speier@ucla.edu}}

\date{}

\renewcommand{\headeright}{}
\renewcommand{\undertitle}{}

\begin{document}
\maketitle

\begin{abstract}
\textbf{Purpose:} The objective of this work is to introduce an advanced framework designed to enhance ultrasound images, especially those captured by portable hand-held devices, which often produce lower quality images due to hardware constraints. Additionally, this framework is uniquely capable of effectively handling non-registered input ultrasound image pairs, addressing a common challenge in medical imaging.

\textbf{Materials and Methods:} In this retrospective study, we utilized an enhanced generative adversarial network (CycleGAN) model for ultrasound image enhancement across five organ systems. Perceptual loss, derived from deep features of pretrained neural networks, is applied to ensure the human-perceptual quality of the enhanced images. These images are compared with paired images acquired from high resolution devices to demonstrate the model's ability to generate realistic high-quality images across organ systems.

\textbf{Results:} Preliminary validation of the framework reveals promising performance metrics. The model generates images that result in a Structural Similarity Index (SSI) score of 0.722, Locally Normalized Cross-Correlation (LNCC) score of 0.902 and 28.802 for the Peak Signal-to-Noise Ratio (PSNR) metric.

\textbf{Conclusion:} This work presents a significant advancement in medical imaging through the development of a CycleGAN model enhanced with Perceptual Loss (PL), effectively bridging the quality gap between ultrasound images from varied devices. By training on paired images, the model not only improves image quality but also ensures the preservation of vital anatomic structural content. This approach may improve equity in access to healthcare by enhancing portable device capabilities, although further validation and optimizations are necessary for broader clinical application.


\end{abstract}


\vspace{1cm}


\section{Introduction}

Ultrasound imaging holds a pivotal role in the field of medical diagnostics, offering non-invasive, highly accurate point of care assessments that have been increasingly adopted by healthcare professionals. Historically, the technology has been limited to large, expensive devices typically found in specialized medical settings. However, there has been a transformative shift towards the development and adoption of compact, hand-held ultrasound devices. These smaller devices promise to democratize access to medical imaging by making it more affordable and widely available. Yet, the miniaturization and cost-effectiveness often come at the expense of image quality, a trade-off primarily attributable to hardware constraints~\cite{Zhou2021}.

Machine learning algorithms have been explored to enhance low-quality images without hardware improvements. For instance, generative adversarial networks (GANs) have been used to create high-quality reconstructions of ultrasound images and videos, offering a cost-efficient avenue for the enhancement of portable ultrasound devices~\cite{Zhou2020,Xia2022,Huang2022}. The CycleGAN framework has become increasingly popular for image-to-image translations, especially those not requiring paired data~\cite{Zhu2017}. The technology has been applied across a spectrum of tasks including, style transfer~\cite{Chang_2018_CVPR} and object transfiguration~\cite{Zhu2017,Ye2019,chen2018attention}. In medical imaging, CycleGANs have been employed in tasks such as pixel-wise translation in echocardiography~\cite{Teng2020}. CycleGANs have also been applied in cross-modality medical image translation such as converting CT to MRI~\cite{Zhang_2018_CVPR}. The architecture has even found utility in histopathology to standardize microscopy staining for more accurate diagnoses~\cite{gadermayr2019generative}.

We hypothesize that the integration of computational algorithms, particularly CycleGAN, can mitigate the disparities in images acquired from different medical imaging devices. Traditional training approaches for these models artificially introduce corruptions into medical images to create pixel-wise pairs~\cite{Geng2018,Chen2017,Eun2020}. However, these methods typically fail to encapsulate the different characteristics of images acquired using different devices. Acquiring paired images using different devices leads to technical issues as images are captured at different time-instances with varying orientations, leading to structural changes that cannot be completely resolved using image registration.

To overcome these challenges, our approach leverages perceptual loss (PL), which can eliminate the need for registration and more accurately relate images from disparate domains. Traditional loss functions used in CycleGAN can result in hallucinated features in the enhanced images~\cite{Cohen2018}. By incorporating PL, more interpretable images are generated that are more robust to registration artifacts~\cite{Armanious2020}. This method can enhance the reliability and consistency of images from handheld ultrasound devices to bridge the gap with expensive high-end systems for greater equity in access to healthcare.

\section{Materials and Methods}

\subsection{Model overview}

\begin{figure*}[!t]
    \centering
    \includegraphics[width=\textwidth]{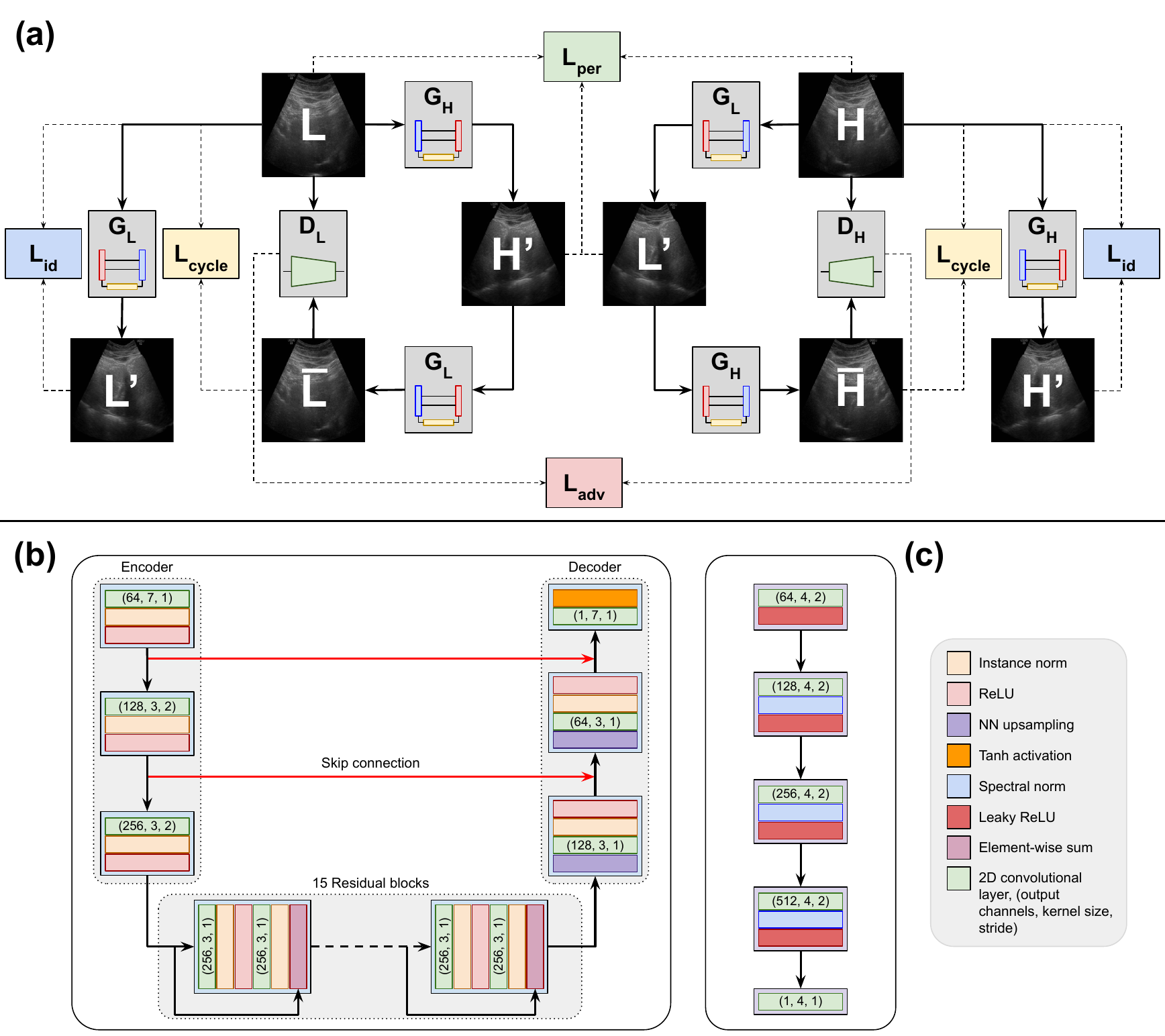}
    \caption{The CycleGAN architecture. (a) An overview of the CycleGAN model training and loss computation framework. (b) The Generator and (c) The Discriminator model architectures. We note: the generators $G_L$ and $G_H$ share the same architecture. Similarly, the discriminators $D_L$ and $D_H$ share the same model architecture. The figure legend lists the different layers in the model.}
    \label{fig:architecture}
\end{figure*}

Our framework for generating high-quality images is a modification of the CycleGAN architecture, designed to map between two distinct imaging domains. In ultrasound image enhancement, these domains correspond to low-quality (Domain L) and high-quality (Domain H) images. The model employs two generators, $G_L$ and $G_H$ , and two discriminators, $D_L$ and $D_H$ (Fig.~\ref{fig:architecture}). The $G_L$ generator transforms an image from Domain H to align with Domain L, whereas $G_H$ performs the inverse, which is crucial for enhancing low-quality images to high-quality images. The discriminators aim to distinguish real images in their respective domains from those transformed by the generators. A unique feature of this approach is the cycle consistency loss that ensures that an image translated to the other domain and then reverted, closely resembles the original. After training, the $G_H$ generator is utilized to enhance images, maintaining essential structural attributes while improving clarity and resolution. 


\subsection{Model description}

GANs have seen transformative advancements, with CycleGAN representing a significant milestone in facilitating unsupervised image-to-image translations. The GAN architecture comprises two primary modules: the Generator and the Discriminator.

The Generator is segmented into encoding, residual block, and decoding phases. The encoder starts by applying $64$ channels of 2D convolution utilizing a large kernel size to capture broader contextual details. Down-sampling layers in the encoder reduce spatial dimensions while increasing depth, thereby emphasizing hierarchical feature extraction. The skip connections in the sequence of 15 residual blocks preserve information to address the vanishing gradient problem.   
Each residual block contains two convolutional sub-layers with a kernel size of 3 and padding of 1. The first layer includes Instance Normalization and a  $ReLU$ activation function, while the second omits the activation function, thereby enabling a linear combination with the block’s input. The Decoder performs spatial restoration and channel reduction using nearest-neighbor interpolation for upsampling and skip connections to preserve detailed features. We use the $tanh$ activation function to scale the output values.

The Discriminator distinguishes between real and generated images. We use spectral normalization to ensure stability during training. Its architecture begins with a convolutional layer that compresses spatial information and expands depth followed by a $LeakyReLU$ activation layer. Subsequent layers maintain the use of spectral normalization to ensure  $1-Lipschitz$ continuity. This constraint on the spectral norm of each layer’s weights helps to balance the generator and discriminator during training. The Discriminator concludes by reducing spatial dimensions to a  $30\times30$ grid with a depth of $1$, providing the final classification output.


\subsection{Perceptual loss}
Conventional methodologies like MSE and SSI rely on pixelwise alignment which makes them unsuitable for non-registered image pairs acquired using different devices. The \textit{Learned Perceptual Image Patch Similarity} (LPIPS) metric addresses these constraints by evaluating the perceptual similarity between images~\cite{zhang2018perceptual}. LPIPS leverages deep features extracted from pre-trained convolutional networks, such as the VGG network~\cite{simonyan2014very}. The LPIPS metric comparing images $X$ and $Y$ is given by:
\begin{equation}
\Phi(X, Y) = \sum_{i} w_i \cdot \| F_i(X) - F_i(Y) \|_2
\end{equation}
where $F_i$ and $w_i$ denotes the feature maps and optimized weights from the $i^{th}$ layer of the pretrained network. Deep feature maps are systematically extracted from every layer within the network, ensuring a comprehensive reflection of the multi-scale characteristics of human perceptual judgment. These features are then unified through linear combination, optimizing the weights to align with perceptual judgments assessed by human evaluators. The LPIPS metric consistently outranks traditional metrics, showcasing superior performance across an array of perceptual judgment tasks~\cite{Armanious2020}. This loss is calculated between real images $L$ and $H$, and those generated through the CycleGAN framework’s generators as:

\begin{equation}
L_{per}(G_H,G_L) = \Phi(H, H') + \Phi(L, L')
\end{equation}

where $H'$ represents $G_H(L)$ and $L'$ represents $G_L(H)$. 


\subsection{Loss function}

\paragraph{Generator loss:} The generator’s loss function is a linear combination of several distinct loss terms, each playing a pivotal role in optimizing image translation between the two domains. First, adversarial loss $L_{adv}(D_H,D_L)$ induces the discriminators to perceive generated images as genuine, whether they are translated from low to high quality or vice versa.  

\begin{equation}
L_{adv}(D_H,D_L) = MSE(\mathbf{1}, D_H(\overline{H})) + MSE(\mathbf{1},D_L(\overline{L}))
\end{equation}

Where $\overline{H}=G_H(G_L(H))$ and $\overline{L}=G_L(G_H(L))$. 
Specifically, mean squared error ($MSE$) calculates the discrepancy between the discriminator’s predictions and an array of ones. These terms push the generator to produce images that can convince the discriminator they belong to the high-quality domain. The cycle loss $L_{cycle}(G_H,G_L)$ prevents the loss of critical image features by ensuring that an image translated to the other domain and back yields the original image. 

\begin{equation}
L_{cycle}(G_H,G_L) = \|H - \overline{H}\|_1 + \|L - \overline{L}\|_1
\end{equation}

The identity loss $L_{id}(G_H,G_L)$ ensures that an image from the source domain fed into the generator remains unchanged.

\begin{equation}
L_{id}(G_H,G_L) = \|H - G_H(H)\|_1 + \|L - G_L(L)\|_1
\end{equation}

Finally, the aggregate generator loss, $L(G_H,G_L,D_H,D_L)$, is computed by combining all individual loss terms weighed by their respective lambda constants:

\begin{equation}
    L(G_H,G_L,D_H,D_L) = \lambda_{adv} L_{adv}(D_H,D_L) + \lambda_{cycle} L_{cycle}(G_H,G_L) + \lambda_{id} L_{id}(G_H,G_L) + \lambda_{per} L_{per}(G_H,G_L) 
\end{equation}

By employing this multifaceted loss function, the model ensures that the generators achieve high-quality image translations while preserving the intrinsic characteristics of the source domain.

\paragraph{Discriminator loss:} The discriminator loss function is designed to evaluate the authenticity of images, incorporating the principle of label smoothing to further enhance the model's generalizability. The discriminator is tasked with distinguishing between real images from the dataset and fake images generated by the corresponding generator.

\begin{equation}
L(D_H,D_L) = 
\begin{cases} 
MSE(\mathbf{1}, D_H) + MSE(\mathbf{1},D_L) & 
\text{if } \text{mean}(D_H) \text{ and } \text{mean}(D_L) < 0.9,\\ 
MSE(\mathbf{0.9}, D_H) + MSE(\mathbf{0.9},D_L) & 
\text{otherwise}.
\end{cases}
\end{equation}

For each domain, the discriminator computes scores for both real and fake images. Conventionally, discriminators are trained using hard labels, where real images are labeled as `$1$' and fake images as `$0$'. However, hard labels can cause vulnerability to adversarial perturbations and lead to overconfidence. Label smoothing addresses this issue by softening the labels; if the mean scores of both real and fake images for the high-quality domain are less than 0.9, a factor of 1.0 is used. Otherwise, a smoothing factor of 0.9 is applied, meaning the real images are given a target value slightly less than 1, which prevents overconfidence and promotes model robustness. The total discriminator loss, $L(D_H,D_L)$, is then computed by aggregating the individual $MSE$ losses $L_{D_H}$ and $L_{D_L}$ for high and low-quality domains respectively.


\subsection{Implementation details}

All models were trained for 300 epochs with a batch size of 24 images. We used the Adam optimizer for model optimization, with separate learning rates set for the discriminators (LR = $3\times 10^{-3}$) and generators (LR = $3\times 10^{-4}$) to implement the two time-scale update rule~\cite{Huang2022}. A beta value of 0.9 for the first and second moments in each optimizer. A learning rate scheduler reduced learning rates by half ($\gamma = 0.5$) every 100 epochs, to allow adaptability during training. Weights were assigned to each loss term: $\lambda_{adv} = 1$ for adversarial loss, $\lambda_{cycle} = 12$ for cycle-consistency loss, $\lambda_{id} = 1$ for identity loss, and $\lambda_{per} = 1$ for PL. The training set was split with 90\% for model training and 10\% for validation. Gradient scaling was used to optimize the model's precision and speed.


\begin{figure*}
\includegraphics[width=\textwidth]{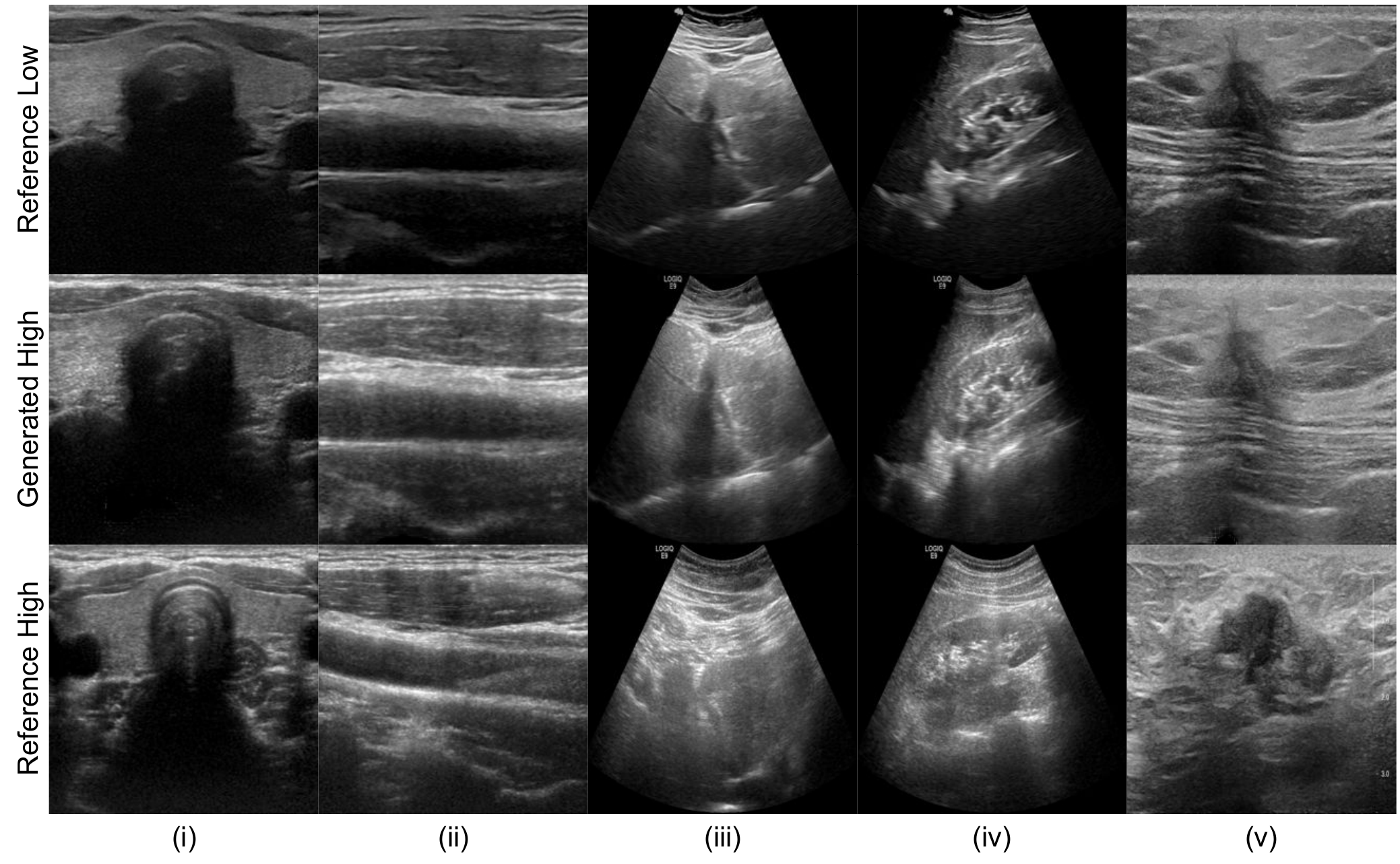}
\caption{High quality images generated in comparison to reference during validation. \textbf{First row:} Reference low-quality images, \textbf{Second row:} Generated high-quality images, \textbf{Third row:} Reference high-quality images. Each column represents scans from (i) Thyroid, (ii) Carotid, (iii) Liver, (iv) Kidney, and (v) Breast.}
\label{fig:val_images}
\end{figure*}

\subsection{Evaluation}

Synthetic high-quality images are evaluated using three metrics: Structural Similarity Index (SSI), Locally Normalized Cross-Correlation (LNCC), and Peak Signal-to-Noise Ratio (PSNR). SSI estimates the structural fidelity between generated and actual high-quality images. LNCC measures the local similarity in intensity patterns between them. PSNR captures the error signal strength derived from the mean squared error (MSE) between $H'$ and $H$.


\begin{table}[!ht]
\captionof{table}{Dataset summary across different ultrasound devices and organs~\cite{USEnhance}.}
\noindent\makebox[\textwidth]{
\begin{tabular}{lllrrrrrr}
\toprule
\multirow{2}{*}{Organ} & \multirow{2}{*}{Low end device} & \multirow{2}{*}{High end device} & \multicolumn{3}{c}{Patients} & \multicolumn{3}{c}{Ultrasound Image Pairs} \\ \cmidrule{4-9} 
& & & Training & Testing & Total & Training & Testing & Total \\ 
\midrule
Thyroid & mSonics MU1 & Toshiba Aplio 500 & 33 & 14 & 47 & 363 & 154 & 517 \\
Carotid & SSUN & Toshiba Aplio 500 & 54 & 23 & 77 & 375 & 160 & 535 \\
Abdomen & SSUN & General Electric LOGIQ E9 & 21 & 9 & 30 & 210 & 90 & 300 \\
\multirow{2}{*}{Breast} & \multirow{2}{*}{mSonics MU1} & Aixplorer ultrasound system & \multirow{2}{*}{23} & \multirow{2}{*}{9} & \multirow{2}{*}{32} & \multirow{2}{*}{102} & \multirow{2}{*}{46} & \multirow{2}{*}{148} \\ 
& & (SuperSonic Imaging S.A.) & & & &  \\
\midrule
\multicolumn{3}{c}{Overall} & 131 & 55 & 186 & 1050 & 450 & 1500 \\ 
\bottomrule
\end{tabular} }
\label{tab:dataset}
\end{table}

\section{Results}

Our study employs a dataset consisting of 3,000 ultrasound images, including 1,500 pairs of low and high-quality images (Table~\ref{tab:dataset}). These images were collected from 186 patients with suspected thyroid tumors, carotid plaque, or breast cancer, along with healthy volunteers. During scans, volunteers were instructed to hold their breath for approximately 10 seconds to minimize deformation, and landmark points were noted for nonrigid registration to ensure the creation of accurate data pairs. This well-curated dataset provides a robust foundation for this study. This dataset was provided by the organizers of the MICCAI 2023 USEnhance challenge~\cite{USEnhance}.

The CycleGAN without PL~\cite{Huang2022} shows improvement over Pix2Pix~\cite{Isola2017} in each metric (Table~\ref{tab:results}). Notably, performance improves when PL is introduced in the CycleGAN models. When increasing the number of residual blocks to 15 the highest performance is observed, with an LNCC of $0.902$, an SSI of $0.722$, and a PSNR of $28.802$. These results underscore our model’s superior capability in retaining both local and global image similarities, ensuring a high-fidelity transformation from low to high-quality ultrasound images while preserving essential diagnostic features (Fig.~\ref{fig:val_images}).

\begin{figure*}[!ht]
\includegraphics[width=\textwidth]{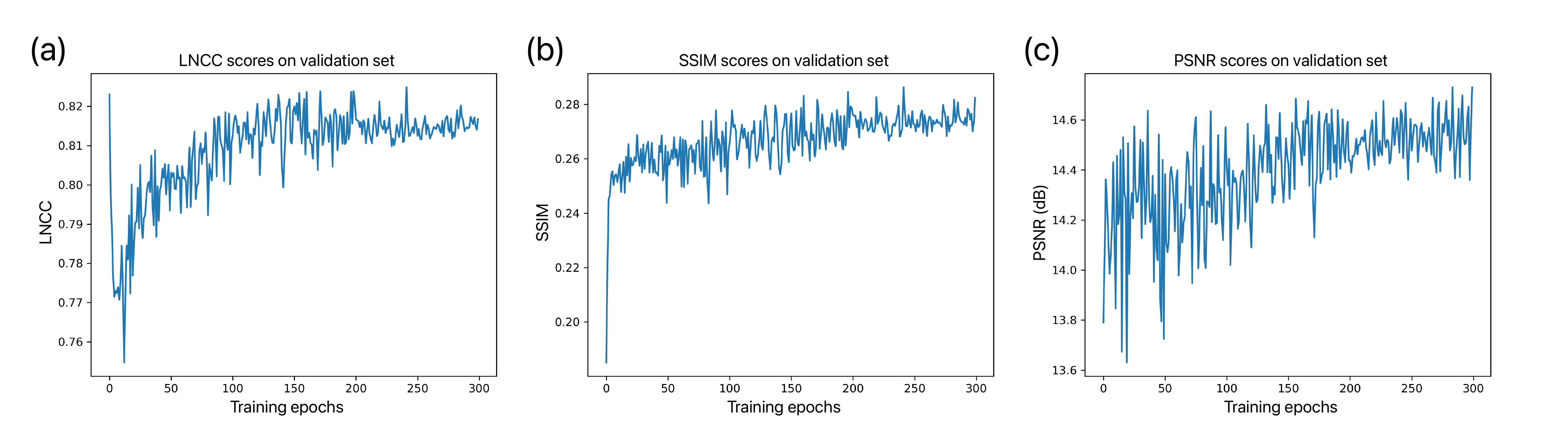}
\caption{Validation set results during model training. \textbf{(a)} Variation of LNCC scores across training epochs, \textbf{(b)} Variation of SSI scores across training epochs, \textbf{(c)} Variation of PSNR scores across training epochs.} \label{fig:val_results}
\end{figure*}

We can notice a stable trend in both the SSI and LNCC scores after 150 epochs of training (Fig.~\ref{fig:val_results}). Although the PSNR scores show some fluctuations, they generally exhibit an upward trend, suggesting the model’s continuous effort to better understand the relationship between low-quality and high-quality domains. 

\begin{table}[!ht]
\centering
\captionof{table}{Test set results.}
\begin{tabular}{lccc}
    \toprule
    Model configurations & LNCC & SSI & PSNR \\
    \midrule
    Pix2Pix                          & 0.669 & 0.343 & 19.732 \\
    CycleGAN (No PL)                & 0.850 & 0.570 & 23.123 \\
    CycleGAN ($\Phi(H,\overline{H})$) & 0.861 & 0.579 & 22.697 \\
    CycleGAN ($\Phi(H,H')$)           & 0.884 & 0.702 & 26.024 \\
    \textbf{Ours}                    & 0.902 & 0.723 & 28.802 \\
    \bottomrule   
\end{tabular}
\label{tab:results}
\end{table}


\section{Discussion}

In this work, we used the CycleGAN framework and PL to bridge the gap between ultrasound images captured by different devices of varying quality. Using PL to compare paired images led to superior results. These findings suggest the possibility of bridging the gap between low and high-quality ultrasound images, a crucial advancement for portable, handheld devices that often struggle with image quality due to hardware limitations. 

Using paired images from different devices allows for a more realistic representation of the image quality differences that would be seen in actual clinical practice. Previous studies have treated the low and high quality domains as independent~\cite{Huang2022}, which can address general image quality issues but does not allow for more direct comparison of image content. The use of PL in this study allows the model to directly compare image from the different domains, improving quality while assuring the preservation of the anatomic content.

One limitation is that the reliance on PL increases computational complexity leading to longer training times. Furthermore, the findings need to be validated through extensive real-world applications and on diverse datasets to ascertain the model's robustness and applicability across different ultrasound imaging scenarios. While this model may work well across organ systems and diseases, future work could also explore whether certain domains would benefit from a more targeted model.

This study introduced an advanced CycleGAN model for ultrasound image enhancement across domains by using PL to train on paired images. These findings demonstrate the feasibility of bridging the image quality gap across devices, leading to improved healthcare equity.


\bibliographystyle{IEEEtran}

\bibliography{references}

\end{document}